\documentclass[12pt]{article}
\usepackage{graphicx}
\usepackage[bookmarksnumbered,colorlinks,plainpages]{hyperref}

\begin{document}

\title{A composition of different $q$ nonextensive systems
with the normalized expectation based on escort probability}

\author{Q.A. Wang, L. Nivanen, A. Le M\'ehaut\'e\\
Institut Sup\'erieur des Mat\'eriaux du Mans, \\
44, Avenue F.A. Bartholdi, 72000 Le Mans, France}

\date{}

\maketitle

\begin{abstract}
This is a study of composition rule and temperature definition for nonextensive
systems containing different $q$ subsystems. The physical meaning of the multiplier
$\beta$ associated with the energy expectation in the optimization of Tsallis
entropy is investigated for the formalism with normalized expectation given by
escort probability. This study is carried out for two possible cases: the case of
the approximation of additive energy; and the case of nonadditive energy prescribed
by an entropy composition rule for different $q$ systems.
\end{abstract}


\section{Introduction}

The extension of the nonextensive statistics (NS) theory developed initially for
systems having same $q$ value in different
formalisms\cite{Tsal88,Curado,Tsal98,Wang01} to the cases of different $q$
subsystems has been a major concern of the scientists in or interested by this
domain\cite{Nauenberg,Tsal03}. The key point in this problematic is the composition
of different $q$ systems into a total system and the interpretation of the
multiplier $\beta$ (measurable temperature) associated with the expectation of
energy in the optimization of Tsallis entropy $S_q$\cite{Tsal88}. Although some
mathematical details are still under investigation\cite{Ou05,Parvan}, the physical
definition of $\beta$ has been satisfactorily clarified for the same $q$ subsystems
(see \cite{Wang04a} for a general review). Essential of $\beta$ interpretation in NS
is to find, for the nonextensive subsystems in equilibrium or stationary states
(with local equilibrium) optimizing the $S_q$ of composite system, the right
quantities that are equal in every subsystems. For example, for thermal equilibrium
or local equilibrium, the temperature $T$ must be qual in every subsystems in order
to be measurable, i.e., $T(A)=T(B)= ...$ (zeroth law) where $A$ and $B$ represent
subsystems. For dynamic and mechanical equilibrium, the pressure $P$ (or equivalent
quantities) must be intensive and equal everywhere, i.e., $P(A)=P(B)=
...$\cite{Wang04a}.

In this work, we limit ourself in the discussion of thermal equilibrium and local
thermal equilibrium (at the point of measure) concerning temperature. As is well
known, for same $q$ subsystems, the starting point is the composition rule of
entropy given by
\begin{equation}                                \label{1}
S_q(A+B)=S_q(A)+S_q(B)+(1-q)S_q(A)S_q(B)
\end{equation}
where the total system $A+B$ is composed of two subsystems $A$ and $B$. However, for
the description of composite systems containing different $q$ subsystems, this rule
should be extended in order to include different $q$'s.

The fundamental reasons why this rule must be extended have been previously
discussed\cite{Wang04a,Wang05a,Wang05b}. For the two formalisms of NS with
unnormalized energy expectation $U_q=\sum_ip_i^qE_i$
($\sum_ip_i=1$)\cite{Wang04a,Wang05a} and $U_q=\sum_ip_iE_i$
($\sum_ip_i^q=1$)\cite{Wang05b}, respectively, a possible extension of Eq.(\ref{1})
has been proposed, i.e.,

\begin{eqnarray}                                    \label{1x}
(1-q_{A+B})S_q(A+B) &=& (1-q_{A})S_{q_A}(A)+(1-q_{B})S_{q_B}(B) \\
\nonumber &+& (1-q_{A})(1-q_{B})S_{q_A}(A)S_{q_B}(B),
\end{eqnarray}
where $q$ is the parameter for $A+B$, $q_A$ for $A$ and $q_B$ for $B$.

In view of the fundamental importance and the wide
application\cite{Abe99,Abe01,Mart01,Mart00,Toral} of the formalism of NS based on
the normalized energy expectation defined with the escort probability
($U_q=\sum_ip_i^qE_i/\sum_ip_i^q$ with $\sum_ip_i=1$)\cite{Tsal98}, it would be
useful to see the possibility of extending Eq.(\ref{1}) to different $q$ subsystems.

This work is a trial of the above approach within this formalism which has an energy
probability distribution given by\cite{Tsal98}

\begin{equation}                                        \label{2}
p_i=\frac{[1-(1-q)\frac{\beta}{\sum_{i}^w p_i^q} (E_i-U_q)]^\frac{1}{1-q}}{Z_q}
\end{equation}
where $\beta$ is the Lagrange multiplier associated with
$U_q=\frac{\sum_ip_i^qE_i}{\sum_ip_i^q}$, $E_i$ is the energy of the state $i$, $w$
the total number of states, the partition function is given by
$Z_q=\sum_{i}^w[1-(1-q)\frac{\beta}{\sum_{i}^w p_i^q} (E_i-U_q)]^\frac{1}{1-q}$ or
$Z_q=\sum_{i}^w[1-(1-q)\frac{\beta}{\sum_{i}^w p_i^q} (E_i-U_q)]^\frac{q}{1-q}$, and
we have $\sum_ip_i^q=Z_q^{1-q}$. For same $q$ nonextensive systems having
Eq.(\ref{1}) and (approximately) additive energy $U_q(A+B)=U_q(A)+U_q(B)$, the
measurable temperature $T$ has been defined by
$\frac{1}{T}=[1+(1-q)S_q]^{-1}\frac{\partial S_q}{\partial
U_q}=Z_q^{q-1}\frac{\partial S_q}{\partial
U_q}=Z_q^{q-1}\beta$\cite{Abe99,Abe01,Mart01,Mart00,Toral} (Boltzmann constant
$k_B=1$).

\section{Different $q$ systems with additive energy}
The formalism of NS with additive energy or other extensive variables is widely
investigated for systems having Tsallis
energy\cite{Abe99,Abe01,Mart01,Mart00,Toral,Vives}. Now let us suppose $A$ and $B$
have respectively $q_A$ and $q_B$ and Eq.(\ref{1x}) holds. When the composite
systems $(A+B)$ optimizing Tsallis entropy, we have
\begin{eqnarray}                                    \label{3}
\frac{(1-q_{A})dS_{q_{A}}(A)}{Z_{q_A}^{1-q_A}(A)}
+\frac{(1-q_{B})dS_{q_{B}}(B)}{Z_{q_B}^{1-q_B}(B)}=0.
\end{eqnarray}
Now using the additive energy rule $U_q(A+B)=U_q(A)+U_q(B)$ and the condition that
the total energy $U_q(A+B)$ is conserved, we get
\begin{eqnarray}                                    \label{4}
\frac{(1-q_{A})}{Z_{q_A}^{1-q_A}(A)}\frac{\partial S_{q_A}(A)}{\partial
U_{q_A}(A)}=\frac{(1-q_{B})}{Z_{q_B}^{1-{q_B}}(B)}\frac{\partial
S_{q_B}(B)}{\partial U_{q_B}(B)},
\end{eqnarray}
which means that at thermal equilibrium or local thermal equilibrium (at the point
of measure) optimizing Tsallis entropy, the intensive quantity that is equal in both
$A$ and $B$ is $\beta'=\frac{(1-q)}{Z_q^{1-q}}\frac{\partial S_q}{\partial
U_q}=\frac{(1-q)\beta}{Z_q^{1-q}}$. So the physical (measurable) temperature $T$
should be defined by
\begin{eqnarray}                                    \label{5}
\frac{1}{T}=\frac{(1-q)}{Z_q^{1-q}}\frac{\partial S_q}{\partial U_q}.
\end{eqnarray}
It is apparent that this definition only hold for $q_A$ and $q_B$ which are both
different from unity. But for the case with one of $q_A$ and $q_B$ equal to one, the
temperature may diverge if $\frac{\partial S_q}{\partial U_q}$ is finite. For the
general validity of NS and the concomitant thermodynamics for different $q$ systems,
this result should be avoided. A possible way for avoiding this is to use
nonadditive energy prescribed by the nonadditivity of entropy Eq.(\ref{1x}) as has
been done for the cases of unnormalized expectation\cite{Wang04a,Wang05a}.

\section{Different $q$ systems with nonadditive energy}

For the purpose of deducing the energy nonadditivity, we take into account the
probability composition rule,
\begin{eqnarray}                                    \label{6}
p_{ij}^q(A+B)=p_i^{q_A}(A)p_i^{q_B}(B)
\end{eqnarray}
produced by Eq.(\ref{1x}). Considering $\sum_ip_i^q=Z_q^{1-q}$, we can write
\begin{eqnarray}                                    \label{6x}
Z_{q}^{1-q}(A+B)=Z_{q_A}^{1-q_A}(A)Z_{q_B}^{1-q_B}(B).
\end{eqnarray}
which can be recast into
\begin{eqnarray}                                    \label{x6}
(1-q)\ln Z_{q}(A+B)=(1-q_A)\ln Z_{q_A}(A)+(1-q_B)\ln Z_{q_B}(B)
\end{eqnarray}
or in differential form:
\begin{eqnarray}   \nonumber                                 \label{6xx}
(1-q)\frac{1}{Z_{q}(A+B)}\frac{\partial Z_{q}(A+B)}{\partial U_{q}(A+B)}dU_{q}(A+B)
\\ = (1-q_A)\frac{1}{Z_{q_A}(A)}\frac{\partial Z_{q_A}(A)}{\partial U_{q_A}(A)}
dU_{q_A}(A)+(1-q_B)\frac{1}{Z_{q_B}(B)}\frac{\partial Z_{q_B}(B)}{\partial
U_{q_B}(B)}dU_{q_B}(B).
\end{eqnarray}
Now from the definition of $Z_q$ given in the introduction, we can find
\begin{eqnarray}                                    \label{xx9x}
\frac{\partial Z_q}{\partial U_q}=Z^q\beta.
\end{eqnarray}
Put this into Eq.(\ref{6xx}) and considering $dU_{q}(A+B)=0$, we see the following
energy nonadditivity:
\begin{eqnarray}                                    \label{9x}
\frac{(1-q_{A})dU_{q_A}(A)}{Z_{q_A}^{1-q_A}}
+\frac{(1-q_{B})dU_{q_B}(B)}{Z_{q_B}^{1-q_B}}=0.
\end{eqnarray}
Comparing Eq.(\ref{9x}) to Eq.(\ref{3}), we obtain
\begin{eqnarray}                                    \label{10x}
\frac{\partial S_{q_A}(A)}{\partial U_{q_A}(A)}=\frac{\partial S_{q_B}(B)}{\partial
U_{q_B}(B)}.
\end{eqnarray}
Considering the relationships $S_q=\frac{Z_q^{1-q}-1}{1-q}$\cite{Tsal98} and
Eq.(\ref{xx9x}), we obtain $\beta=\frac{\partial S_q}{\partial U_q}$. So the
Lagrange multiplier $\beta$ in the distribution Eq.(\ref{2}) should be taken as the
measurable inverse temperature. Hence the definition of physical temperature with
normalized energy expectation given by escort probability is the same as with the
two unnormalized expectations mentioned above\cite{Wang05a,Wang05b} if we take into
account the nonadditivity of energy prescribed by the entropy nonadditivity.

\section{Conclusion}
We have studied the composition of different $q$ nonextensive systems and the
definition of temperature for this kind of systems with normalized expectation of
energy given by the escort probability. Essential of our approach is to find the
quantity that is equal in all the subsystems at thermal equilibrium or stationary
states optimizing Tsallis entropy. The starting point of this work is a
nonadditivity rule of entropy Eq.(\ref{1x}) which generalized the rule of
Eq.(\ref{1}) for same $q$ systems. The conclusion of this work is that, if we
consider the nonadditivity of energy prescribed by the nonadditivity of entropy, the
measurable (physical) temperature is give by $\frac{1}{T}=\frac{\partial
S_q}{\partial U_q}$. This result is the same as obtained with the two unnormalized
expectations of energy in our previous work\cite{Wang04a,Wang05a,Wang05b}.

Summarizing the works on these three possible formalisms of NS, the temperature
definition is the same as in the conventional statistical mechanics. As a
consequence, the form of the first and second laws of thermodynamics does not change
in the nonextensive thermodynamics associated with these formalisms. However, it is
not the case with other normalized expectations used in the case of same $q$
subsystems where the physical temperature is given by $\frac{\partial S_q}{\partial
U_q}$ multiplied by a function of the partition function $Z_q$\cite{Wang04a}. To our
opinion, these formalisms with normalized expectations are not suitable for
different $q$ systems described by Eq.(\ref{1x}).

\section*{Acknowledgement}
The authors would like to thank Dr. A. EL Kaabouchi for fruitful discussions.

\end{document}